\documentclass{article}
\usepackage{spconf,amsmath,graphicx}
\usepackage{cite}
\usepackage{amsmath,amssymb,amsfonts}
\usepackage{algorithmic}
\usepackage{graphicx}
\usepackage{textcomp}
\usepackage{xcolor}
\usepackage{siunitx}
\usepackage{svg}
\usepackage{subfig}

\usepackage[shortcuts,acronym]{glossaries}
\usepackage{makecell}
\usepackage{multirow}

\usepackage{caption}
\usepackage{booktabs}
\usepackage{makecell}
\usepackage{multirow}

\usepackage{todonotes}

\newacronym{fmcw}{FMCW}{frequency modulated continuous wave}
\newacronym{rdi}{RDI}{range-Doppler image}
\newacronym{fft}{FFT}{fast Fourier transformation}
\newacronym{snr}{SNR}{signal-to-noise ratio}
\newacronym{relu}{ReLU}{rectified linear unit}
\newacronym{mflops}{MFLOPs}{mega floating point operations}

\title{Light-weight Gesture Sensing Using FMCW Radar Time Series Data}
\name{Thomas Stadelmayer$^{1,2}$, Avik Santra$^{2}$, Robert Weigel$^{1}$, Fabian Lurz$^{3}$}%\thanks{Thanks to XYZ agency for funding.}}
\address{$^{1}$Friedrich-Alexander University Erlangen-Nuremberg, $^{2}$Infineon Technologies AG,\\ $^{3}$Hamburg University of Technology}
\begin{document}
%\ninept
%
\maketitle
\begin{abstract}
The paper proposes a novel feature extraction approach for FMCW radar systems in the field of short-range gesture sensing. A light-weight processing is proposed which reduces a series of 3D radar data cubes to four 1D time signals containing information about range, azimuth angle, elevation angle and magnitude. The processing is entirely performed in the time domain without using any Fourier transformation and enables the training of a deep neural network directly on the raw time domain data. It is shown experimentally on real world data, that the proposed processing retains the same expressive power as conventional radar processing to range-, Doppler- and angle-spectrograms. Further, the computational complexity is significantly reduced which makes it perfectly suitable for embedded devices. The system is able to recognize ten different gestures with an accuracy of about 95\% and is running in real time on a Raspberry Pi 3 B. The delay between end of gesture and prediction is only \SI{150}{\milli\second}.

%Radar offers a promising modality for enabling dynamic gesture recognition for human-machine interface. In this paper we propose a novel feature extraction approach wherein a set of 1D time series data is directly obtained from the raw ADC data, thus avoiding the need for range-Doppler processing. We demonstrate that these time series data retain the same expressive power as feature images such as classical range, Doppler and angle spectrograms, while the signal dimension is reduced from 2D to 1D. Consequently, not only the computational complexity for preprocessing is significantly reduced, but also the computational complexity of further processing and classification is reduced by an order of magnitude and thus providing a perfectly suited system for embedded devices. Furthermore, we propose a novel deep representation learning approach where class centers in the embedding space are learned. Similar to triplet loss a push and pull mechanism is used to design discriminative class clusters. In contrary to triplet loss our proposed solution can be trained in a end-to-end fashion and without the need of triplet mining, while the classification performance is competitive. We demonstrate our proposed gesture sensing solution using a 60-GHz radar and show the improved performance of our proposed classifier to triplet loss and cross-entropy loss.

\end{abstract}

\section{Introduction}
Radar-based gesture recognition systems provide users an intuitive human-machine interface as an alternative to traditional click and touch based interfaces. Gesture recognition has several applications ranging from smart TVs, laptops, smart phones to controlling robotics. Compared to camera-based solution, radar-based gesture solutions are privacy preserving, and can sense minute subtle gesture motions \cite{santra2020deep, lien2016soli}. 

%Typically radar-based gesture classification is a two step procedure. First, the raw radar data is preprocessed to range or Doppler spectrograms \cite{DS_hc_ml, doppler_spectro, rd_based} or sequence of range-Doppler and/or range-angle images \cite{souvik_rdm, continuous_rdm}. Then the classification is performed based on hand crafted features extracted from the signals, mainly in combination with machine learning, or by interpreting the preprocessed spectrograms or series of range-Doppler images (\ac{rdi}s) by deep neural networks.

Multiple papers have already proven the capability of mm-wave radar sensors to recognize various gestures with high accuracy often enabled by interpreting preprocessed data with machine or deep learning \cite{DS_hc_ml, doppler_spectro_cnn, profiles_lstm_cnn, souvik_rdm_3dcnn, time_data_cnn, rdm_3dcnn}. The challenges nowadays are to make the systems robust against distortions such as persons walking in the background \cite{multiperson} and to reduce the system complexity in order to enable the market of embedded devices like smart phones \cite{radarnet_google_cnn_lstm}.
%For the former challenge the range selectivity of frequency modulated continuous wave (\ac{fmcw}) radars provides a great advantage over continuous wave (CW) radar solutions. For the latter challenge more efficient processing and data representation in combination with advanced neural network architectures is needed.

In fact, recent papers are more and more focusing on the implementation on embedded or edge devices. In \cite{mateusz_2dcnn} a 2D CNN optimized for Intel's Neural Compute Stick 2 is doing the classification based on range spectrograms. However, the angle of arrival is not estimated and thus only four basic gestures can be recognized.
In \cite{sun_yuliang_2dcnn} a feature cube is created by selecting K points from each \ac{rdi} and estimating their range, Doppler, azimuth angle, elevation angle and magnitude. The feature cubes, which are a compressed representation of the gestures, are interpreted by a 2D CNN. The system is running in real time on a NVIDIA Jetson Nano.
The TinyRadarNN presented in \cite{tinyRadarNN_tcn} and the RadarNet \cite{radarnet_google_cnn_lstm} from the ATAP Google team follow a similar principle. The data is preprocessed to \ac{rdi}s, which are then image wise compressed to a feature vector by a 2D CNN. A series of compressed \ac{rdi}s is then interpreted by a temporal neural network. In contrast to many other papers the authors of \cite{radarnet_google_cnn_lstm} highlight the importance of the phase information and thus are working with complex valued \ac{rdi}s.%A drawback of this approach is the non-interpretability of the compressed feature vectors.

%and the larger amount of training data needed.
%In \cite{deterministic_signal_ml} a continuous wave radar with three receiving antennas ordered as a L-shape is used. The classification is done based on the 1D phase displacement signal of the three antennas. However, only macro gestures, i.e. entire hand movements, and no micro gestures defined by single finger movements can be recognized. Further due to missing range selectivity random body movements might effect the system.

%proposes to compress the \ac{rdi}s by a 2D CNN and further process the compressed signal by a temporal convolutional network. The gestures are solely classified by the range and Doppler information and thus making it impossible to distinguish basic gestures as e.g. swipe from left to right and swipe from right to left, which require an angle estimate.
%In \cite{radarnet_google_cnn_lstm} the ATAP Google team presents their solution called RadarNet. Unlike many other papers working on the absolute values of the \ac{rdi}s and thus removing information, the authors propose to work on the complex complex valued \ac{rdi}s. Each \ac{rdi} is then compressed by a neural network to only 32 values called frame summary. Then a temporal network interprets a series of frame summaries. However, also this works relies on range-Doppler processing by a 2D FFT. Further, the compression of the \ac{rdi}s is done by a neural network, which leads to abstract and non-interpretable frame summaries.

All current solutions using a \ac{fmcw} radar rely on \ac{rdi}s, which are generated by a 2D \ac{fft} with high computational effort.
%The \ac{rdi}s are either compressed by classical processing techniques to spectrograms or a feature cube or by a neural network to feature vectors.
To the best of our knowledge this is the first work providing an alternative to the dominating \ac{fft} based preprocessing for \ac{fmcw} radars.
By minimal processing -- without any \ac{fft} -- a series of 3D radar data cubes is converted into relative range, accumulated azimuth as well as elevation angle and magnitude over slow time. %Therefore, the signal represents movements in all possible directions.
Using real world data we demonstrate the classification performance of a deep neural network trained by our proposed time series data in comparison with conventional range-, Doppler, azimuth and elevation spectrograms. The system is implemented on a Raspberry Pi 3 B and working in real time. The contributions of the paper are as follows:
\vspace{-0.1cm}
\begin{itemize}
    \itemsep-0.25em 
    \item A novel feature extraction algorithm is presented as alternative to conventional range-Doppler processing.
    %\item Direct interpretation of slow time data by a neural network is enabled.
    \item The competitiveness of the proposed processing compared to conventional \ac{fft} based processing is shown on real world data.
    %\item The system works in real time on a Raspberry Pi 3.
    \item Low complexity of the proposed processing is shown.
\end{itemize}

\section{Gesture Set and Radar System Design}
\label{sec:dataset}
\begin{figure}[htbp]
\centerline{\includegraphics[width=0.43\textwidth]{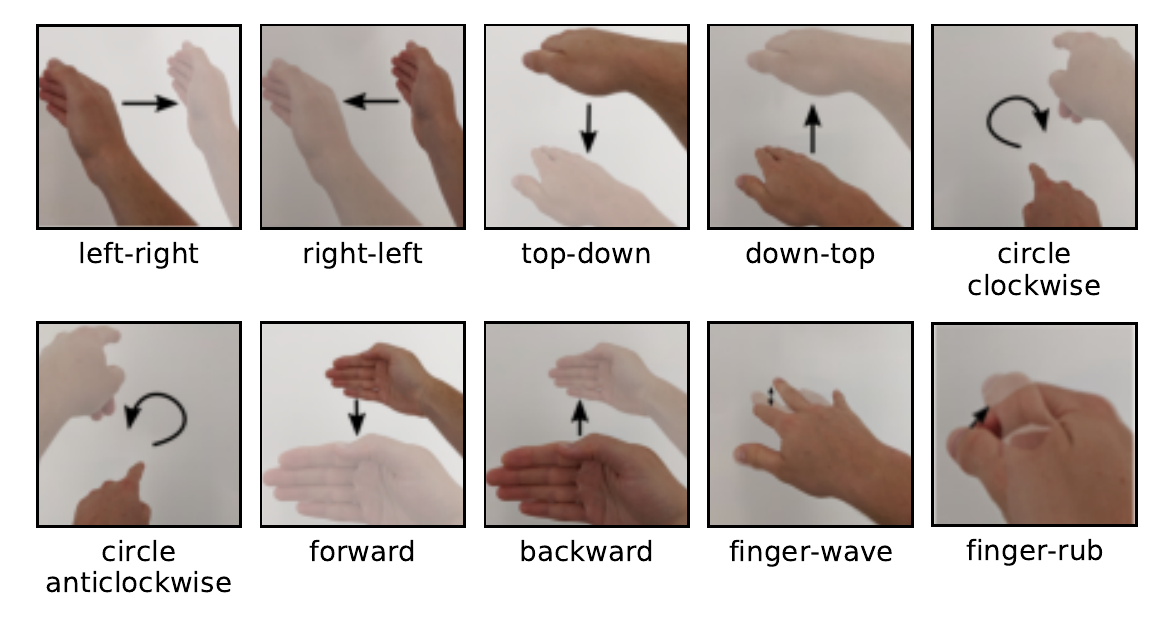}}
\caption{Set of gestures.}
\label{fig:gestures}
\end{figure}

The set of gestures used in the paper includes eight macro and two micro gestures. A macro gesture is a motion of the entire hand such as swipe left to right, whereas a micro gesture is defined by single finger motions such as rubbing with the thumb over the index finger. In Fig.~\ref{fig:gestures} the set of gestures is depicted.

The gesture recognition system presented in the paper uses Infineon's 60-GHz \ac{fmcw} radar chipset BGT60TR13C for data acquisition. The radar chipset has one transmit and three receive antennas. The receivers are ordered in a L-shape, which enables an angle estimation on the basis of two antennas each in azimuth as well as elevation direction.\\
The radar is configured to send chirps ranging from $f_{\text{min}}$ = \SI{58}{\giga\hertz} to $f_{\text{max}}$ = \SI{63}{\giga\hertz}. 
With a pulse repetition time of $T_{\text{PRT}}$ = \SI{0.39}{\milli\second} the radar sends a burst of 32 chirps. Each received chirp is mixed with the transmit signal. The resulting intermediate frequency signal then undergoes an anti aliasing filter and is sampled 64 times with a frequency of \SI{2}{\mega\hertz}. The signal of such a set of chirps is called data frame and is immediately transferred to the host. The data frames are produced in a frequency of \SI{30}{frames\per\second}.

The received and digitized signal is stored in a 4D array $s_\text{0}[f, r, n, m]$ with the dimensions F$\times$R$\times$N$\times$M, where $f$ is the data frame index, $r$ the receive channel index, $n$ the slow-time index and $m$ the fast-time index. The signal contains undesired DC components, mainly arising from antenna leakage and static objects in the field of view, which have to be removed. Therefore, for each data frame and for each receiving channel the mean is removed along fast-time and along slow-time dimension. The mean removed signal $s[f, r, n, m]$ is the basis for any further processing introduced in this paper.

\section{Conventional Signal Processing}

The frequency of the IF signal depends on the time delay of the received signal and thus is coupled on the range of the back scattering object. Thus, conventionally a \ac{fft} along each chirp is applied to obtain the range information within the signal. Motion of the back scattering object result in a change of phase within the corresponding fast-time frequency or range bin introducing the Doppler shift. Since the phase is periodical, the frequency within a range bin across multiple chirps indicates the radial speed of the target.
%Thus, it is very common to process the data frames using a 2D \ac{fft}, which provides then a \ac{rdi}. Mathematically, the \ac{rdi} is defined as 
%\begin{align}
%&\textrm{v}_{\text{\ac{rdi}}}(p,l) =
%&\bigg{|} \sum_{m=1}^{N_{\text{st}}} \sum_{n=1}^{N_{\text{ft}}} w[m,n] \text{s}[m,n] e^{-j2\pi %(\frac{mp}{N_{\text{st}}} + \frac{nl}{N_{\text{ft}}}) }  \bigg{|} \notag \\
%\end{align}
%where $w(n,m)$ is a 2D window function and $\text{s}(n,m)$ is the signal of an arbitrary data frame and a receiving channel. The $n, m$ are the indices along fast-time and slow-time, while $l, p$ are the indices along range and Doppler. $N_{\text{st}}$ and $N_{\text{ft}}$ define the number of chirps and number of samples per chirp respectively.

\textit{Range and Doppler spectrograms:} To obtain a real valued signal, the amplitude of the \ac{rdi}s is used in this processing step. Since only a single target within the field of view can be assumed in short range gesture sensing the 2D \ac{rdi}s can be marginalized in both dimensions to get a range and a Doppler vector, without losing much information about the object of interest, whereas the data dimension can be reduced. The range and Doppler vectors of consecutive frames are concatenated and form a 2D range or Doppler spectrogram respectively. The range and Doppler spectrograms of the three receiving channels are finally integrated to increase the \ac{snr}.

%For the range spectrogram the range vector with maximum signal energy is selected. Correspondingly, the Doppler vector with maximum signal energy over all Doppler frequencies is selected and appended to the Doppler spectrogram.

\begin{figure}[htbp]
\centering
\includegraphics[width=0.4\textwidth]{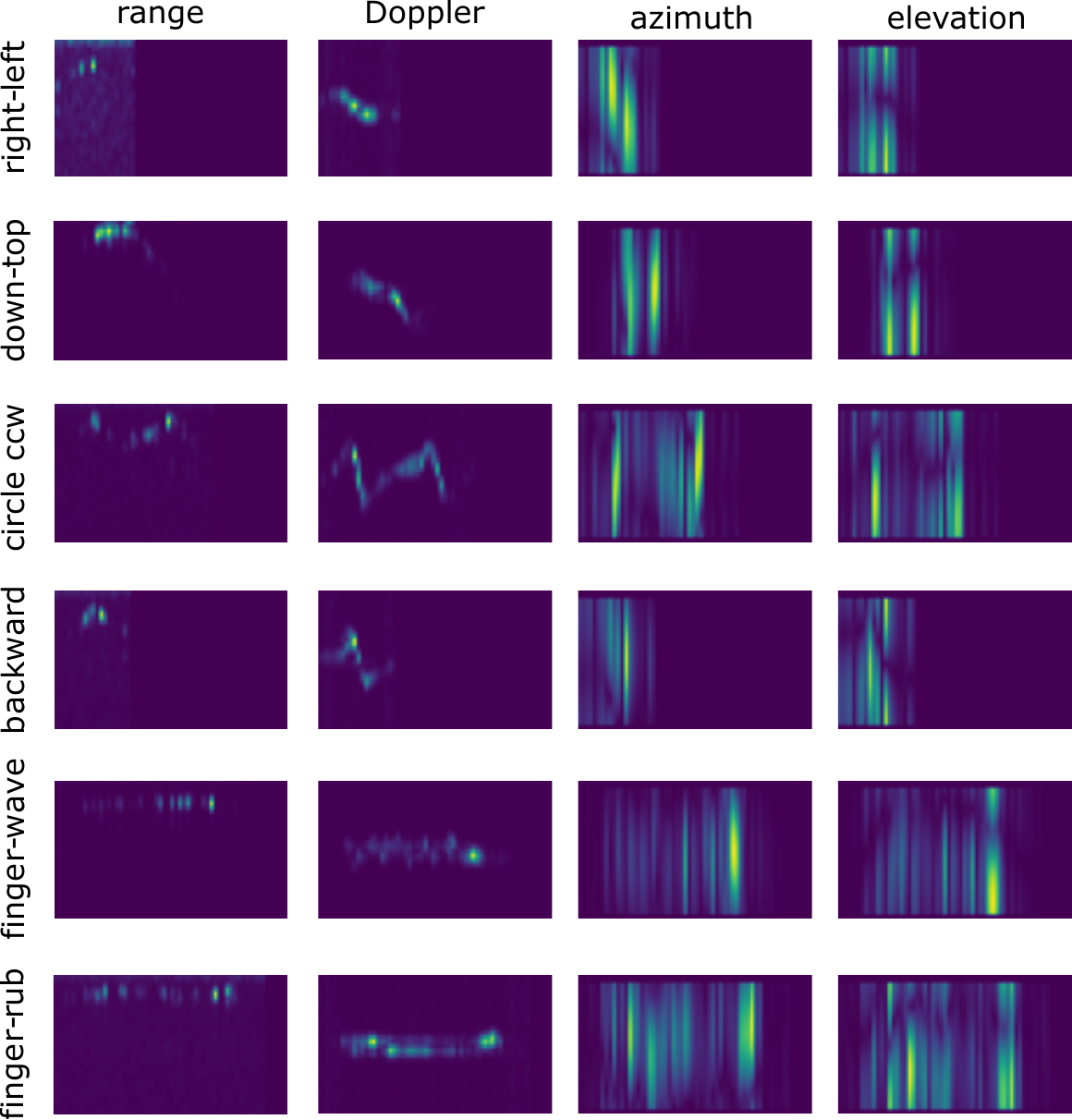}
\caption{Exemplary set of spectrograms for selected gestures. The x-axis of the spectrograms is the time dimension going from 0 to \SI{2}{s}.}
\label{fig:spectro_gestures}
\end{figure}

\textit{Angle spectrogram:} The angle of arrival is estimated based on the range-Doppler bin with the highest amplitude in the \ac{rdi}. Based on the complex value of the range-Doppler bin found by a peak search on the amplitudes a digital beamforming expressed as 
\begin{eqnarray}
a(\hat{\theta}) = \sum_{r \in R} x_r \exp\left(-j\frac{2\pi d^r \sin(\hat{\theta})}{\lambda}\right)
\label{eq:dbf}
\end{eqnarray}
is applied. In (\ref{eq:dbf}) $R$ is the set of receiving channels across which the digital beamforming should be applied, $x_r$ is the complex valued selected range-Doppler bin of the $r^{\text{th}}$ channel and $\hat{\theta}$ is the estimated angle sweeped across the field of view at predefined angular steps. For azimuth angle estimation $R$ is the set of antenna 1 and 3 and for elevation angle estimation $R$ is the set of antenna 2 and 3. Exemplary images of a set of range, Doppler, azimuth and elevation spectrograms are depicted in Fig.~\ref{fig:spectro_gestures}.

%\iffalse
%\fi

%\section{Proposed Solution}
%In this paper we present two novel approaches. First, the conventional 2D \ac{fft} based preprocessing is replaced by an efficient processing procedure resulting in a set of 1D signals. The processing steps and its advantages are discussed in the first part of the section. In the second part we describe our proposed classifier, that is inspired by D-Softmax and triplet loss, while it overcomes difficulties like triplet mining and allows an end-to-end training. 

\section{Proposed Time Series Processing}
The used radar chipset provides a real valued signal. To obtain the motion and angle information the phase of the signal is needed. To achieve this, in conventional processing a \ac{fft} is applied along fast time. The \ac{fft} divides the signal into N range bins with a range resolution of $\frac{c}{2BW}$, where $BW$ is the chirp bandwidth. As also underlined by \cite{lien2016soli}, the fundamental sensing paradigm is not the range profile of the hand, since the range resolution is anyway too coarse, but the analysis of the scattering center dynamics through processing of temporal changes that occur in the raw signal over slow time. Therefore, it is superfluous to resolve the signal in range direction using a \ac{fft}, which comes with a high computational complexity. Instead, we propose to apply a complex valued low pass sinc filter to the fast time signal. This has two positive effects: First, the low pass filter acts as a limitation of the field of view, thus masking out all motions outside of the region of interest and second, the signal is transformed to complex domain which allows an easy access to the phase information with high temporal resolution.

For low pass filtering a complex valued 1D sinc filter is used. The filter is defined as
\begin{eqnarray}
&h[m, f_{\text{c}}, b] = 2b~\text{sinc}(2bt) e^{jmT}
\end{eqnarray}
%\begin{eqnarray}
%&h[m, f_{\text{c}}, b] =\nonumber\\
%&2b~\text{sinc}(2bt) \left(\cos(f_\text{c} mT) + j \sin(f_\text{c} mT)\right)
%\end{eqnarray}
with $m \in \left[-\lfloor\frac{M}{2}\rfloor, \lfloor\frac{M}{2}\rfloor\right]$, where $M$ is the filter length, $T$ the sampling interval, $f_\text{c}$ the center frequency and $b$ the bandwidth. Since the signal is static within a chirp, no convolution per se is performed but the fast time signal is just multiplied by the complex valued sinc filter. Thus, the signal within a data frame after filtering is complex valued and one dimensional. The signal is defined as
\begin{eqnarray}
s[f, r, n] = s[f, r, n, m] \cdot h[m, f_{\text{c}}, b] \cdot w[m]
\end{eqnarray}
where $w[m]$ is a window function. Further, $f_{\text{c}}$ is chosen to be 0.25 normalized frequency and $b$ is set to 0.5 normalized frequency. The resulting complex valued signal is stored as magnitude $m[f, r, n] = |s[f, r, n]|$ and phase $\phi[f, r, n] = arg(s[f, r, n])$.

\textit{Magnitude:}
The magnitude array is flattened by concatenating the signal of multiple frames followed by an integration over all receiving channels. As a result the array holding the magnitude of the signal is simply defined as $m[k]$, where $k$ ranges from 0 to F$\times$N.

\textit{Azimuth and elevation angle:}
To extract information about the angle of arrival the phase difference between two antennas $r_0$ and $r_1$ is used as defined in (\ref{eq:phase_diff}). 
\begin{eqnarray}
\Delta\phi_{\text{$r_0$, $r_1$}}[f, n] = \phi[f, r_0, n] - \phi[0, r_1, n]
\label{eq:phase_diff}
\end{eqnarray}
The relation between phase difference in spatial dimension and the angle of arrival in the 1D case can be expressed as $\theta = \sin^{-1}{(\frac{\lambda \Delta\phi}{2\pi d})}$ where $\lambda$ is the wavelength and $d$ the antenna distance. For the used radar the antenna distance $d$ is half a wavelength. Given this fact, the relation between phase offset and angle of arrival can be approximated to be linear. Thus, for simplicity we directly interpret the phase difference as angle of arrival.

Also the phase difference signal is flattened over the frames similar as it was done for the magnitude signals. Thus, the phase difference signal becomes the 1D array $\Delta\theta_{\text{$r_0$, $r_1$}}[k]$. 
To increase the \ac{snr} the spatial phase difference is accumulated. Therefore, the resulting signal representing information about the angle of arrival is defined as
\begin{eqnarray}
\text{aoa}_{\text{$r_0$, $r_1$}}[k] = \sum_{i}^{k} \Delta\phi_{\text{$r_0$, $r_1$}}[i].
\end{eqnarray}
Note that due to integration the steepness of the signal represents the angle of arrival. The array containing information about the azimuth angle is defined by choosing antenna 1 and 3 as $r_0$ and $r_1$ whereas the elevation angle is estimated between antenna 2 and 3. Therefore, the azimuth array is defined as $\text{az}[k] = \text{aoa}_{\text{1, 3}}[k]$ and the elevation array is defined as $\text{el}[k] = \text{aoa}_{\text{1, 2}}[k]$.
\iffalse
\begin{eqnarray}
\text{az}[k] = \text{aoa}_{\text{1, 3}}[k]
\end{eqnarray}
and the elevation array is defined as
\begin{eqnarray}
\text{el}[k] = \text{aoa}_{\text{1, 2}}[k].
\end{eqnarray}
\fi

%and the phase signal is two dimensional $\Phi[r, k]$, where $r$ is the receiving channel index and $k$ the chirp index.
%and flattened to a 1D array per channel. 
\textit{Relative range:}
By phase unwrapping in slow time direction the relative range can be estimated. However, the idle time between data frames cause a random phase offset between first sampling point in the current frame and the last sample in the previous frame. Therefore, simply concatenating the frames and then evaluating the phase displacement would lead to discontinuities between the frames. Instead the phase difference in slow time direction is first estimated within each frame $\Delta\phi[f, r, n] = \phi[f, r, n+1] - \phi[f, r, n]$. Then the phase difference signals are concatenated over frame dimension. Due to the difference the resulting signal is F data points smaller then the magnitude and angle arrays. Thus between the frames a phase difference is approximated by a linear interpolation between the last and first phase difference of two consecutive frames as defined in (\ref{eq:phase_diff}). The resulting signal is then integrated over the receiving channels, which finally leads to another 1D signal of size F$\times$N describing the targets relative radial position over time.

Thus, in total the radar signal of a recorded gesture is represented as four 1D time series' namely the relative range, the integrated azimuth phase offset, the integrated elevation phase offset and the absolute value. An illustrative example of a set of range, azimuth, elevation and absolute value time series for each gesture is plotted in Fig.~\ref{fig:time_gestures}. Despite the idle time between the frames the 1D time series data is smooth and without discontinuities. Thus, it can be nicely processed by a 1D CNN without causing artifacts in the convolution. %\textcolor{red}{We can perfectly reconstruct the raw IQ signal $\rightarrow$ no information is lost in this processing $\rightarrow$ proof??}

%There are multiple advantages of using the time series representation as described above. First, the signal was transformed from a 2D and frame wise signal into a 1D frame continuous signal. Thus, a continuous 1D convolution with individual kernel sizes can be applied without being restricted to the data frames. Second, the approach makes use of the fact that only a single target is within the field of view. In conventional processing a \ac{fft} resolves the entire range Doppler domain, however, when there is only a single target, most of the range Doppler bins will just be zero. With the presented data representation only the target signal is captured. Third, due to integration the SNR of the phase, azimuth and elevation signal is significantly improved. And forth, the target with the highest radar cross section, which is usually the palm of the hand for gesture sensing, leads to the lowest frequency part in the phase signal. Finger movements introduce phase offsets on top of this. Thus, by band pass filtering the macro motion of the hand and the micro motions of the fingers can be independently analysed.
\begin{figure}[htbp]
\centering
\includegraphics[width=0.4\textwidth]{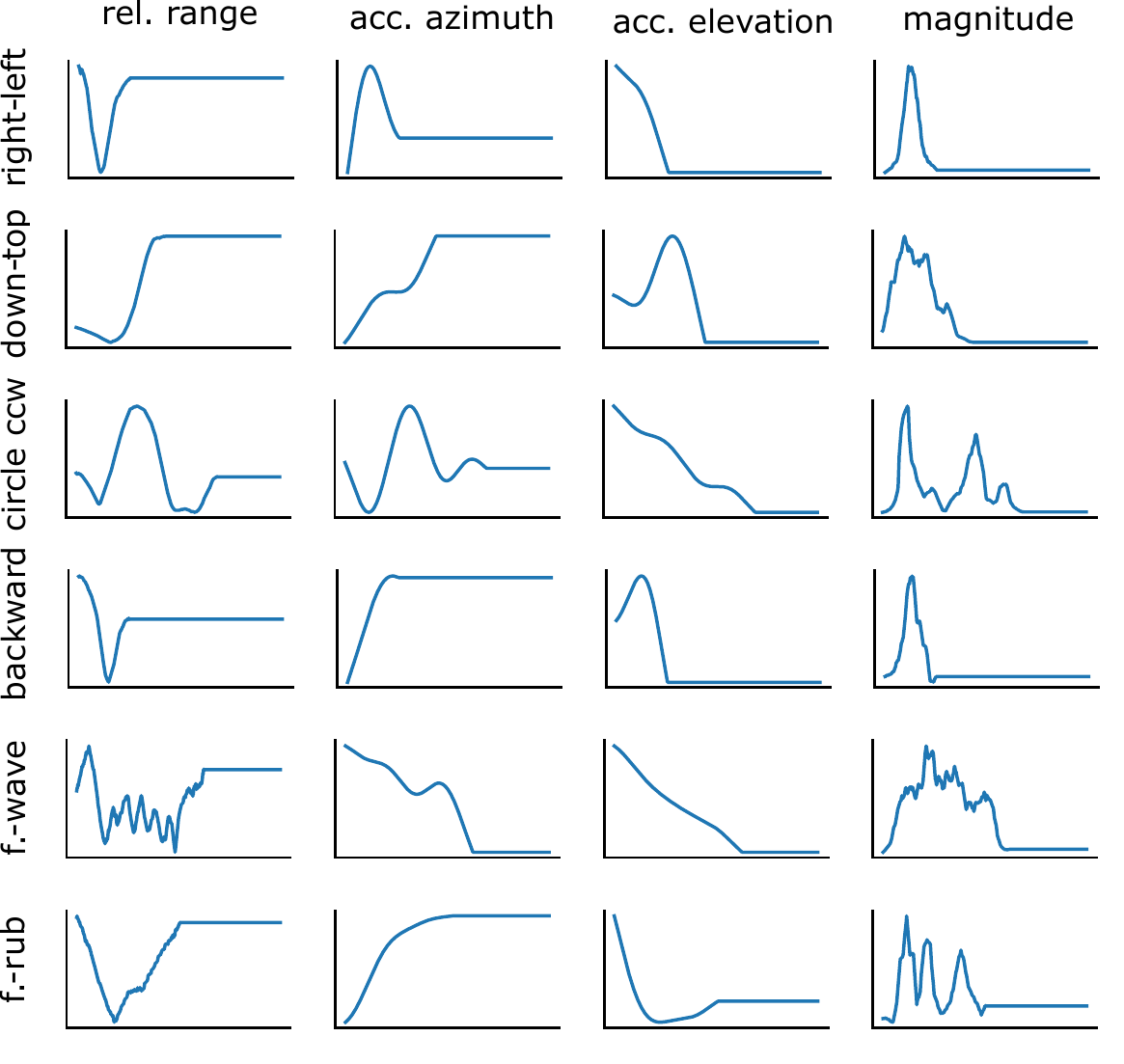}
\caption{Exemplary set of time series for selected gestures. The x-axis of the plots is the time dimension going from 0 to \SI{2}{s}.}
\label{fig:time_gestures}
\end{figure}

\iffalse
\begin{figure}[htbp]
\centering
%\centerline{\includegraphics[width=0.25\textwidth]{figures/circle.png}}
\hspace{-0.5cm}
\subfloat[Left right]{\includegraphics{figures/left_right_labels.pdf}}
\hspace{-0.35cm}
\subfloat[Down top]{\includegraphics{figures/down top.pdf}}
\hspace{-0.35cm}
\subfloat[Circle cw]{\includegraphics{Conference-LaTeX-template_IEEE_SENSORS_2021/figures/ccw.pdf}}
\hspace{-0.35cm}
\subfloat[Forward]{\includegraphics{Conference-LaTeX-template_IEEE_SENSORS_2021/figures/forward.pdf}}
\hspace{-0.35cm}
\subfloat[Finger wave]{\includegraphics{Conference-LaTeX-template_IEEE_SENSORS_2021/figures/finger wave.pdf}}\\
\vspace{0.5cm}

\hspace{-0.5cm}
\subfloat[Right left]{\includegraphics{Conference-LaTeX-template_IEEE_SENSORS_2021/figures/right left.pdf}}
\hspace{-0.35cm}
\subfloat[Top down]{\includegraphics{Conference-LaTeX-template_IEEE_SENSORS_2021/figures/top down.pdf}}
\hspace{-0.35cm}
\subfloat[Circle ccw]{\includegraphics{Conference-LaTeX-template_IEEE_SENSORS_2021/figures/cccw.pdf}}
\hspace{-0.35cm}
\subfloat[Backward]{\includegraphics{Conference-LaTeX-template_IEEE_SENSORS_2021/figures/backward.pdf}}
\hspace{-0.35cm}
\subfloat[Finger rub]{\includegraphics{Conference-LaTeX-template_IEEE_SENSORS_2021/figures/finger rub.pdf}}\\
\caption{Exemplary set of time series for each gesture class.}
\label{fig:time_gestures}
\end{figure}
\fi

\section{Experiment and Results}
In the first part of this section the classification accuracy of two similar ConvNets trained by conventional spectrograms and the proposed time series data, respectively, is compared, whereas the second part discusses the computational complexity of both systems.
The used dataset consists of the 10 gestures introduced in sec.~\ref{sec:dataset}. In total 300 repetitions per gesture were recorded from 8 different people. The train and test dataset where split by person. Thus, 250 samples from 6 persons for each class is used for training and 50 samples from 2 different persons per class is used for testing. Further, each training is repeated 10 times to reduce the variance in the results. 
%\textcolor{red}{start and end of gesture detection??}

\subsection{Classification Accuracy}
For time series classification a 3 layered 1D CNN is used. The first convolutional layer uses 32 filters with a kernel size of 64, the second and third layer are using 32 kernels with a filter width of 16. As very first input layer a average pooling of size 4 is used and after each convolutional layer an max-pooling of size 2 is performed and a \ac{relu} is applied as activation function. After the convolutional block the tensor is flattened and fed into a fully connected layer with 32 output dimensions followed by a fully connected layer with 10 output dimensions and softmax activation.
%Based on the output of this layer the classification is performed. For the softmax and cross-entropy classifier a further fully connected layer with 10 output dimensions is used, for the Euclidean classifier a Euclidean distance layer with 10 output dimensions is used and the triplet loss works directly on the embedded feature vectors.\\
The neural network processing the set of spectrograms is 3 layered 2D CNN with 32, 64 and 64 filters of the sizes 5x5, 3x3 and 3x3. Max pooling is used after each convolutional layer and a \ac{relu} is used as activation. Also here the convolutional block is followed by a fully connected layer with 32 outputs and a fully connected layer with 10 outputs and softmax activation.

Cross entropy is used as loss function and stochastic gradient descent with a learning rate of 0.005 is used as optimizer for both models. The trainings are carried out 10 times and the mean accuracy as well as the standard deviation is used as metric. The classification results along with the number of parameters and the number of \ac{mflops} needed for one prediction are given in Tab.~\ref{tab:result_data}. Both models achieve a similar accuracy of about 96\%.

\subsection{Computational Complexity}
Processing a data frame of size N $\times$ M by a \ac{fft} comes with a computational complexity of $\mathcal{O}(NM \log(NM))$. With the proposed processing the 2D \ac{fft} is replaced by a multiplication of each chirp with a single complex valued sinc filter which has a complexity of $\mathcal{O}(NM)$. Creating the range, azimuth, elevation and magnitude signal is done in $\mathcal{O}(N)$. Thus, the computational complexitiy of processing a data frame was reduced from $\mathcal{O}(NM \log(NM))$ to only $\mathcal{O}(NM)$. It is noteworthy that with \ac{fft} processing the targets angle has to be additionally estimated, whereas the angle information is already extracted in the proposed processing.

Although the ConvNet interpreting the time series data is only one dimensional, it has more parameters than the 2D ConvNet. The reason is that the system has to interpret time series data. Therefore, there are no clear edges as in the image like 2D spectrogram representation. Thus, we are using larger filter sizes to cover a larger receptive field. However, the 1D ConvNet has less then half the \ac{mflops} to perform compared to the 2D ConvNet doing a single prediction.
%Please note that the FLOPs only refer to the operations within the neural network. The preprocessing operations are not even taken into account. Additional to the improved compactness and computational advantages of using the 1D time series representation, the classification accuracy is very similar to the one of using 2D spectrograms.
Thus the hardware requirements are significantly reduced by maintaining the same classification accuracy.

In order to demonstrate the functionality and real-time capability of the system under limited hardware resources, the system was run on a Raspberry Pi 3 B. Neither the preprocessing nor the neural network were optimized. Nevertheless, the preprocessing takes about \SI{100}{\milli\second} and predicting a class takes about \SI{50}{\milli\second} resulting in a reaction time of only \SI{150}{\milli\second} in total.

%Although the time series data representation is more compact and the network smaller, it achieves an even better classification accuracy than using the spectrograms. The neural network is able to directly learn from the raw data and thus find superior features as the manually designed spectrograms can provide. 
%Triplet loss is used as loss function. After training the mean embedded feature vector of each class is used as class center and the final classification is determined by the closest class center.

\begin{table}[htb]
\centering
\scriptsize{
\caption{Comparison of 2D ConvNet using 2D preprocessed spectrograms and 1D ConvNet using the 1D time series as input with respect to classification accuracy, memory footprint and computational complexity.}
\begin{tabular}{@{}cccc@{}}
\toprule
\textbf{Model}     & \textbf{Acc. ($\pm$ dev.)}  & \textbf{\# parameter}  & \textbf{\ac{mflops}} \\ \midrule
\makecell{2D ConvNet\\(spectrograms)}   & \makecell{95.7\%\\($\pm$ 0.9\%)}     & 51488 & 8.960 \\ \midrule
\makecell{1D ConvNet\\(time series)} & \makecell{96.0\%\\($\pm$ 0.8\%)} & 70784  & 3.719 \\ %\midrule         

\midrule
\bottomrule

\end{tabular}
\label{tab:result_data}
\vspace{-4mm}
}

\end{table}

\section{Conclusion}
In this paper a promising alternative to the dominating \ac{fft} based preprocessing for \ac{fmcw} radars was proposed. The proposed processing reduces the computational complexity for preprocessing a single data frame to $\mathcal{O}(NM)$. It was experimentally shown on real world data that the novel processing is capable of achieving comparable classification performance to conventional FFT preprocessed data. 

%approach wherein a set of 1D time series data is directly extracted from raw ADC data, thus completely eliminating the need for conventional range-Doppler processing using 2D fast Fourier transforms. To the best of our knowledge, this is the first paper that proposes a deep learning based classification directly from raw time domain data without range or Doppler processing. %Furthermore, we proposed a novel deep representation learning approach dubbed Euclidean classifier, which learns the intra-class compactness and inter-class separation in the Euclidean space using end-to-end training with a single neural network. Using a 60-GHz short-range radar, we demonstrate a gesture sensing solution using these time series data and proposed Euclidean classifier and compare our results against state-of-the-art representational learning approaches such as triplet loss and D-Softmax.

\vfill\pagebreak

\bibliographystyle{IEEE}
\bibliography{literature}

\end{document}